\def\year{2022}\relax
\documentclass[letterpaper]{article} 
\usepackage{aaai22}  
\usepackage{times}  
\usepackage{helvet}  
\usepackage{courier}  
\usepackage[hyphens]{url}  
\usepackage{graphicx} 
\usepackage{natbib}  
\usepackage{caption} 
\DeclareCaptionStyle{ruled}{labelfont=normalfont,labelsep=colon,strut=off} 
\nocopyright 
\usepackage{xcolor}

\usepackage{amssymb}

\newcommand{\galen}[1]{}
\newcommand{\tim}[1]{}
\newcommand{\amy}[1]{}
\newcommand{\new}[1]{#1}

\newcommand{\appendixlink}{\url{https://arxiv.org/abs/2109.05152}}
\newcommand{\appendixref}[1]{\href{https://arxiv.org/abs/2109.05152}{Online Appendix}~\ref{#1}}

\newcommand\blfootnote[1]{%
  \begingroup
  \renewcommand\thefootnote{}\footnote{#1}%
  \addtocounter{footnote}{-1}%
  \endgroup
}

\usepackage{xspace}
\newcommand{\hide}[1]{}
\newcommand{\xhdr}[1]{\vspace{1.7mm}\noindent{{\bf #1.}}} 

\newcommand{\eg}{\textit{e.g.,}\xspace}
\newcommand{\ie}{\textit{i.e.,}\xspace}
\newcommand{\sect}{\S}

\usepackage{caption}
\usepackage{subcaption}

\usepackage[hidelinks]{hyperref} 

\usepackage{multirow}
\usepackage{tabularx}

\title{Making Online Communities `Better': A Taxonomy \\ of Community Values on Reddit}
\author{
Galen Weld, Amy X. Zhang, Tim Althoff \\
\normalsize{{\normalfont Paul G. Allen School of Computer Science \& Engineering, University of Washington}} \\
\normalsize{\normalfont \{gweld, axz, althoff\}@cs.washington.edu }
}

\begin{document}

\maketitle

\begin{abstract}
Many researchers studying online communities seek to make them better.
However, beyond a small set of widely-held values, such as combating misinformation and abuse, determining what `better’ means can be challenging, as community members may disagree, values may be in conflict, and different communities may have differing preferences as a whole.
In this work, we present the first study that elicits values directly from members across a diverse set of communities.
We survey 212 members of 627 unique subreddits and ask them to describe their values for their communities in their own words.
Through iterative categorization of 1,481 responses, we develop and validate a comprehensive taxonomy of \new{community values, consisting of} 29 \new{subcategories within nine top-level categories}, enabling principled, quantitative study of community values by researchers.
Using our taxonomy, we reframe existing research problems, such as managing influxes of new members, as tensions between different values, and we identify understudied values, such as \new{those regarding} content quality and community size.
We call for greater attention to vulnerable community members' values, and we make our codebook public for use in future research.
\end{abstract}

\section{Introduction}\label{sec:intro}
Online communities account for an ever-increasing share of all human interaction~\cite{williamson_2020_social_media}.
People use millions of different online social communities for accessing news~\cite{weld_political_2021}, for entertainment, and for socialization, amongst many other purposes. Unfortunately, these online communities have been shown to have substantial harms to both community members and society as a whole, including the distribution of  misinformation, harassment and bullying, and coordinated activity to undermine elections. As a result, many researchers are working on methods to understand these harms and make online communities `better.' 
However, truly understanding how to make communities `better' requires going beyond simply mitigating and minimizing harms and going towards an understanding of community \textit{values}.
Determining a community's values is a non-trivial challenge, as communities have many stakeholders with divergent preferences. 
\blfootnote{Additional materials can be found in our Online Appendix at \appendixlink.}

\new{
The term `values' can have many meanings; here we use the definition from Value Sensitive Design: `what a person or group of people consider important [to their online community]' \cite[pg.2]{friedman_2006_valuesensitive}.
In this definition, values not only have a \emph{topic}, \eg the diversity of the community, but also a \emph{preference}, \eg a preference for \textit{more} diversity.
Value Sensitive Design is a design framework which underpins our work.
At its core, Value Sensitive Design suggests that values must be considered when designing in contexts such as online communities \cite{friedman_2006_valuesensitive}.
}

\new{
Understanding community values is challenging because values can vary widely both between and within communities, and values can conflict with one another.
A community focused on mental health support may want to foster inclusion more than a community focused on financial trading, a difference in values between communities.
Within the same community, members may have different value preferences regarding the same value topic, \eg one community member may desire more diversity, while another may wish the community was more homogeneous.
Someone who is a member of two (or more) communities may even have different preferences in different communities; that same person may prefer more diversity for one community and more homogeneity for the other.
Lastly, values of different topics may conflict with one another as well.
For example, while an online photography discussion community may desire to create a space that is welcoming to beginners, this value conflicts with the same community's desire to hear particularly from expert photographers who may be perceived as having the most to contribute to the conversation.
These differences and conflicts are a critical consideration for researchers as well as community moderators and members.
}


Although much has been published on positive aspects of online communities~\cite{robert_e_kraut_building_2012},
previous work \new{which seeks to make online communities better} has largely focused on specific aspects of online communities, especially those which are widely agreed to be harmful, such as harassment, rule-breaking, and misinformation.
However, upon deeper inquiry, even these more commonly studied harms are quite complex, with substantial disagreement within and across communities regarding the extent to which these ostensibly harmful behaviors should be tolerated~\cite{jiang_2021_international_harm, scheuerman_2021_framework_severity}.

Implicit in much of this prior work is an assumption of communities' \new{and their members'} values, yet \new{exactly what values communities hold has not yet been comprehensively studied.}
While we do not argue that any one set of values is superior to others, we believe that a critical first step towards improving online communities is developing a comprehensive understanding of community members' own values, not just obvious ones that researchers have assumed all communities care about.

In this work, we survey redditors to answer the research question ``What values do community members hold for their communities?''
Through a series of advertisements placed on reddit, we recruit 212 people and collect 1,481 free response answers to questions about the values they hold for the 627 unique communities they consider themselves a part of (\sect\ref{sec:method}).
To the best of our knowledge, this is the first such survey to gather community members' values in their own words.



We apply an iterative categorization methodology to produce our primary contribution: a taxonomy of community values with nine top-level categories and 29 subcategories (Fig.~\ref{fig:taxonomy}). We validate this taxonomy with a held-out set of 1,180 additional responses to demonstrate saturation, and achieve very high inter-rater agreement (Fleiss kappa $= 0.874$) using a codebook which we make public to support future research (\appendixref{app:codebook}).
Among the other key findings we contribute, we find that online social community members' value\new{s cover} a broad range of \new{categories} from technical features to the diversity of community members, and that the quality of content submitted to communities is the most frequently reported value \new{category} (\sect\ref{sec:taxonomy}).

Our work enables new research to improve online communities (\sect\ref{sec:discussion}).
In addition to working on what they consider valuable, researchers seeking to improve online communities should work on values we found to be frequently \new{held} yet understudied. Open research challenges with regard to understudied values include measuring the quality of content in the specific context of different communities, and the difficulty of growing communities' membership while maintaining meaningful community interaction (\sect\ref{sec:understudied_values}). 
Now that we have delineated a taxonomy of values, we can also begin to examine how values relate to each other. We argue that some widely held community values are inherently in tension, such as maintaining the quality of content in a community while simultaneously including new members, and call for research into community design with these tensions in mind (\sect\ref{sec:conflicting_values}). Finally, we call for additional work on reconciling differences in values within communities, including community governance that protects the values of vulnerable community members (\sect\ref{sec:governance_discussion}).
\section{Survey and Categorization Methodology}\label{sec:method}

\subsection{Reddit Background \& Context}\label{sec:reddit}
In this work, we focus on reddit. 
reddit is an ideal platform for conducting research on the values of online communities, as (unlike other social media platforms such as Twitter) reddit is explicitly divided into thousands of unique communities, known as `subreddits,' each with their own topical foci, rules, moderators, norms, and enforcement practices. Furthermore, almost all content on reddit is publicly available~\cite{baumgartner_pushshift_2020}, and reddit has been widely examined by the research community~\cite{medvedev_anatomy_2019}.
On reddit, subreddit's names are prefixed with \texttt{/r/}, and we adopt this notation in the rest of this paper.

\subsection{Survey Instrument}\label{sec:instrument}
All responses were gathered through an online survey hosted on the Qualtrics platform. The survey is summarized here and included in its entirety in \appendixref{app:survey}.
The survey consists of five sections: (1) informed consent, (2) demographic questions, (3) general questions about the respondent's usage of reddit, (4) subreddit specific questions, and (5) reflection questions. All questions except the informed consent question are optional.

Before any other questions are posed, the participant is shown a brief summary of the nature and aims of the survey, along with IRB information (for more details, see \sect\ref{sec:ethics}) and is asked for their consent. Then, in the demographics section, the participant is asked to describe their age, gender identity, and racial identity. Due to the challenge of obtaining parental consent over the internet, minors (people under the age of 18 in our jurisdiction) are excluded from participating.

Next, the participant is asked to optionally provide their reddit username, which is used to query the reddit API for a list of subreddits in which the participant recently posted or commented, and as a contact point for the raffle (\sect\ref{sec:incentives}).

Once the general reddit questions are answered, the participant is shown a list of the five most recent subreddits they posted or commented in (\appendixref{app:extra_figs}), and asked to remove any subreddits that they do not consider themselves a member of, and to add any subreddits they consider themselves a member of that are missing from the list. Participants who decline to provide their username or whose username was not found on reddit are presented with an empty list for them to populate themselves, and entered subreddits are checked in real time against the reddit API to preclude spelling errors and to ensure the subreddit exists.

The next section is dedicated to subreddit specific questions. It consists of two free-response questions which are asked in turn for every subreddit listed by the participant in the previous step: \textit{As it exists right now}, what are a few of the best aspects of the /r/\textless subreddit\textgreater community? \textit{If you could change anything}, what are some aspects of the /r/\textless subreddit\textgreater community you would like to improve upon?
\new{
We chose to use these two questions to elicit participants' values from their specific, real experiences, rather than asking participants to speculate about abstract cases.
To this goal, we chose to ask about positive aspects of the community as well as negative aspects, as these two questions elicit different values: the positive question offers insight into participants' values which are well-implemented by their communities, whereas responses to the negative question show which of the participants' values are less accommodated by current community practices.
Asking both of these questions allows us to gather the broadest possible set of values for categorization, crucial to developing a comprehensive understanding of community values.
}

The survey was piloted with 13 participants from several departments in two large American universities. All 13 pilot participants denied having any difficulty understanding any of the survey questions or interacting with the online survey tool.

\subsection{Participant Recruiting and Incentives}\label{sec:incentives}
The 212 participants in this study were recruited primarily through reddit advertisements. These advertisements display inline with other content in both individual subreddits and aggregated content views, and are shown on the reddit website as well as the official reddit mobile app. \appendixref{app:extra_figs} Figure~\ref{fig:recruitment} shows the appearance of these advertisements.
To increase the diversity of recruiting, the survey was also distributed to relevant university mailing lists and Slack channels at two large American universities, as well as posted to /r/SampleSize, a subreddit for the distribution of surveys, recruiting an additional 41 participants.
Participation was incentivized with a raffle.
Additional details on recruiting \new{pipeline} attrition and the raffle are located in \appendixref{app:recruiting}.

\subsection{Participant Demographics}\label{sec:demographics}
In general, we find that the demographics of our 212 survey respondents match overall platform demographics closely~\cite{pew_2016_reddit_demographics} (summary of respondent demographics in \appendixref{app:demographics}).
As with reddit's overall userbase, our respondents skew young, white, and more commonly identify as men. Compared to reddit's overall demographics, the age our of respondents is similar, with 71\% of our respondents being under the age of thirty. People who identify as non-binary or do not identify with man or woman are slightly overrepresented in our survey results compared to reddit as a whole.
When it comes to racial and ethnic identities, white and Black users are slightly underrepresented in our responses, while Hispanic and Asian responses are slightly overrepresented.

\subsection{Iterative Categorization}\label{sec:categorization}
Once the survey responses were collected, all free-text responses were divided into idea units~\cite{strauss_qualitative_1987}, where each idea unit represents a distinct thought. For example, the response `The content is educational and I like how the community is engaged' would be divided into two idea units: `The content is educational' and `I like how the community is engaged.'


\new{Our} taxonomy was produced using an initial set of 301 idea units gathered from the first 39 respondents to the survey, with the remaining 1,180 \new{idea units} held out for validation.
Using a grounded theory approach~\cite{glaser_1968_grounded}, a team of five researchers worked together to iteratively categorize the initial idea units using an inductive coding method~\cite{MacQueen1998CodebookDF}.
\new{
While respondents often expressed value preferences in their responses, only value topics were used to categorize idea units.
}
The researchers worked independently to initially cluster similar idea units, then came together to resolve differences in clustering until a consensus was reached.
The initial tentative clusters were assigned names and definitions to produce a working taxonomy, then the researchers collaboratively recategorized all idea units, creating and removing categories under group consensus.
This process was repeated \new{three} additional times until the iterative process converged and no further changes were needed to satisfactorily categorize all idea units.
Once this was completed, the researchers worked together to write a codebook (\appendixref{app:codebook})
describing the taxonomy, which is hierarchical, with top-level categories and subcategories.
When possible, idea units are assigned to the more specific subcategories, with top-level categories reserved for broad or vague idea units.

\subsection{Inter-Rater Reliability and Validation of the Taxonomy}\label{sec:validation}
To validate the codebook, three researchers were trained on the codebook and independently labeled 100 idea units randomly sampled from the held-out set. Inter-rater reliability was very high, with a Fleiss kappa of 0.874 when considering all 29 subcategories. When measuring agreement on only the nine top-level categories, the three raters were in even greater agreement (Fleiss kappa = 0.902).
This level of agreement would typically be described as ``almost perfect'' \cite{landis_koch_1977_measurement} or ``excellent'' \cite{gwet_2014_irr_handbook}, and demonstrates that the taxonomy categorization using our codebook is robust and largely unambiguous.

To validate the taxonomy, the \new{1,180} held-out idea units were coded by a single researcher. Every single one of the idea units in the validation set was able to be categorized using the taxonomy codebook, demonstrating saturation.

\subsection{Ethical Considerations \& \new{Broader} Impacts}\label{sec:ethics}
In order to ensure the anonymity of our participants, we keep all their responses confidential. Access to responses was limited to only the immediate research team, and all responses quoted in the following sections have been paraphrased to ensure participant anonymity.
All data collected in this study was provided by participants who were informed of the nature of the study, the potential risks of participation, and who consented to participate.
We do not make any use of participants' public reddit histories for any purpose, as the use of such data has been shown to make many participants uncomfortable~\cite{Fiesler2018ParticipantPO}.
This research was reviewed and approved by the University of Washington IRB under ID STUDY00011457

\begin{figure*}[t]
    \centering
    \includegraphics[trim={0 111mm 0 0},clip,width=1.00\textwidth]{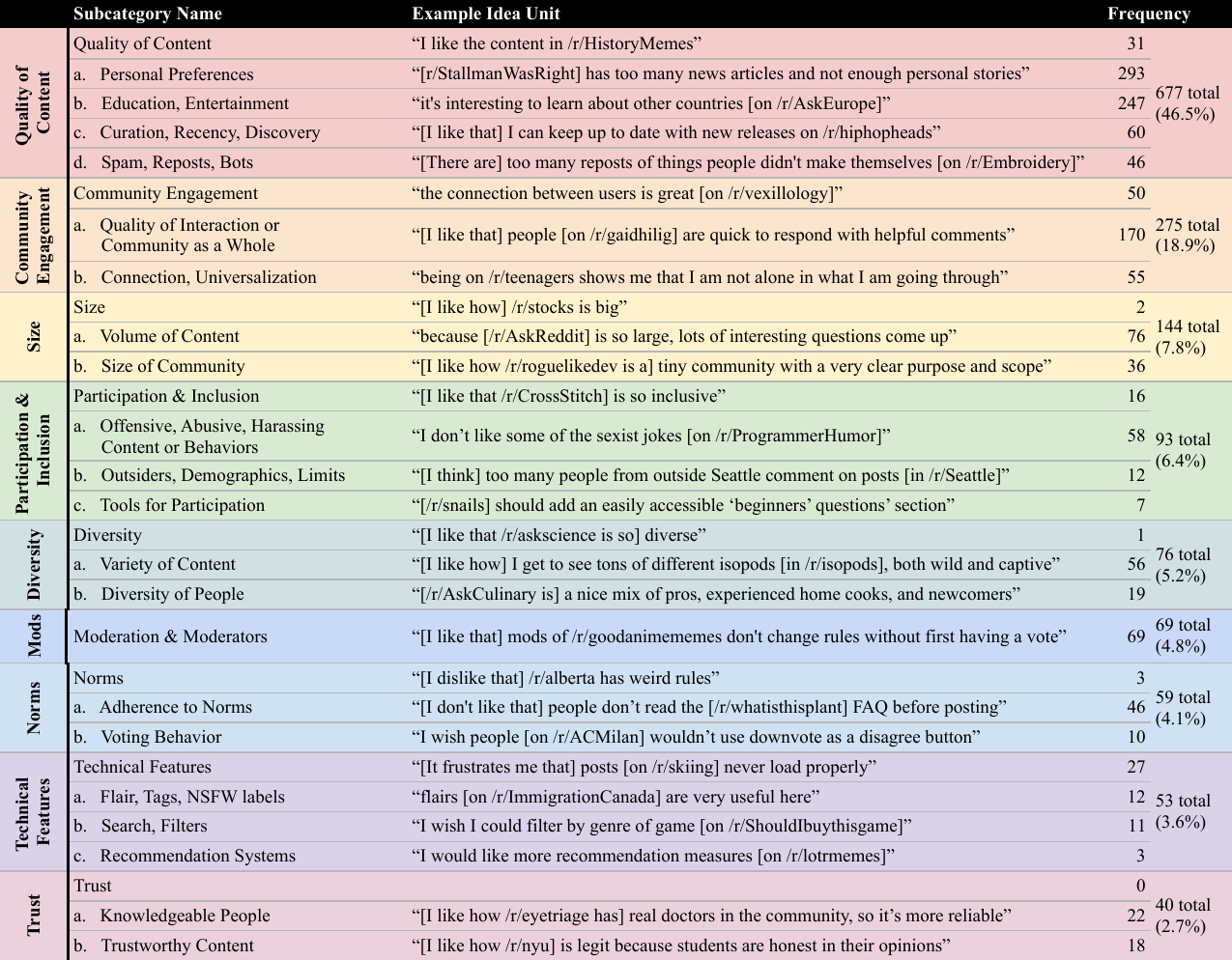}
    \vspace{-5mm}
    \caption{\small Summary of the taxonomy of values produced by iterative categorization of idea units from responses, sorted by frequency. Example quotes from participants are given, along with the total number of idea units for each subcategory and top-level category. On the right, the percentages given in parenthesis give the percentage of all idea units in each top-level category. The community value most frequently commented on is \textit{Quality of Content}, with 46.5\% of all idea units.
    \new{ While we report the frequency with which values were reported in our survey, we note that this does not necessarily mean that frequent values are necessarily the most important ones to consider, nor that this implies agreement on value preferences across and even within communities (\sect\ref{sec:governance_discussion})}
    }
    \vspace{-5mm}
    \label{fig:taxonomy}
\end{figure*}

\section{Taxonomy of Community Values}\label{sec:taxonomy}
Our taxonomy \new{has two hierarchical levels---the lower level} consisting of 29 specific \new{subcategories} grouped into nine top-level categories. These \new{categories reflect the diverse range of value topics reported by our participants, ranging} from \textit{Technical Features} to \textit{Diversity} (Fig.~\ref{fig:taxonomy}).
\new{Categories are grouped according to value topic, regardless of value preference, and thus may contain conflicting value preferences, \eg the Size category contains idea units from respondents who prefer both larger and smaller sizes for their respective communities.}
\textit{Quality of Content} is the \new{most frequently mentioned value category}, with 47\% of all idea units, while \textit{Technical Features} and \textit{Trust} are the least frequently \new{mentioned categories}, each with less than 4\% of the idea units.

\new{
While we report counts for the number of idea units falling into each category and subcategory, we do not make quantitative claims or comparisons between different categories or different communities. Communities, each with their own subject matter, rules, and membership, are different from one another, which is why our survey is designed to ask participants about each community individually.  
Rather than quantify the importance of any one category, 
our goal and contribution is to understand the broad diversity of value topics across communities in greater detail.
}
The following subsections describe each \new{taxonomy category} in detail.

\subsection{Quality of Content}\label{val:quality}
The \textit{Quality of Content} category is the largest of the categories that compose our taxonomy, containing 4 subcategories and 46.5\% (677/1481) of all idea units derived from responses. It is unsurprising that most community members would \new{have many values relating to} the content of the communities they are a part of, as in most reddit communities, sharing content is the primary mode of interaction with the community. We expect \new{values in this category} to be especially \new{prevalent} in communities that are less focused on engagement and connection with others and most focused on content itself (\eg meme sharing communities). 

The four subcategories in this category (\textit{Personal Preferences}, \textit{Education/Entertainment}, \textit{Curation/Recency/Discovery}, and \textit{Spam/Reposts/Bots}) reflect the broad range of aspects of content and its presentation that are \new{liked} or disliked by community members. Many community members \new{appreciate} the curation and discovery of content that is provided by the community through mechanisms such as up- and down-voting, \eg ``I keep up to date with new releases on /r/hiphopheads.'' The most frequently mentioned \new{subcategory within} the top-level \textit{Quality of Content} category is \textit{Personal Preferences}, which reflects how well (or poorly) the content aligns with personal preferences for specific types of content, such as memes (``the best part of /r/Terraria is the memes'') or different \new{subjects} (``[in /r/BalticStates,] the posts about Estonia are the best because it's such a fabulous country'').
These \textit{Personal Preferences} are difficult to generalize because in most cases they are specific to the \new{subject} of the community as a whole.

Idea units regarding \textit{Education/Entertainment} and \textit{Spam/Reposts/Bots} tend to be more consistent from subreddit to subreddit. In general, content that is educational or entertaining is especially \new{liked}, \eg ``/r/knitting teaches me new stitches and patterns'', while \textit{Spam/Reposts/Bots} are fairly universally disliked, and manifest more similarly in different communities. However, the exact nature of reposts and repeated content can vary somewhat from subreddit to subreddit, \eg ``I wish users [of /r/Askreddit] would quit asking the same questions over and over just to get karma\footnote{Karma is reddit's name for points gathered through the receipt of upvotes.}.''

\subsection{Community Engagement \& Interaction}\label{val:engagement}
The \textit{Community Engagement} category contains 18.9\% (275/1481) of the idea units derived from responses, making it the second most frequently reported value \new{topic} amongst respondents.
We divide these idea units into two subcategories: \textit{Quality of Interaction or of the Community as a Whole}, and \textit{Connection and Universalization}, \ie the realization that others exist with similar interests/feelings. The vast majority (256/275) of these responses were mentioned in a positive context, suggesting that redditors mostly feel positively about community engagement.

Comments on \textit{Quality of Interaction or of the Community as a Whole} mentioned both qualities of the individual interactions with community members (\eg ``I often ask for help with language learning resources [on /r/gaidhlig], and everyone is always quick to respond'') as well as qualities of the community as a whole
(\eg ``I appreciate the goodhearted nature of most of the people [on /r/Konosuba, a community dedicated to the eponymous Japanese novel series].'')
However, 19 participants were unhappy with the quality of the interactions in their communities, such as ``[On /r/msu] there are often gatekeepers who comment on posts. However, these people tend to get downvoted quickly, so it's not that big of a deal.'' This community member then suggested that ``a brief note about negativity in the sub's rules could help with this.''

\textit{Connection and Universalization} is especially \new{oft-mentioned} in communities for people with a common identity or interest who may be physically far apart or part of a minority group and therefore unable to connect as easily offline. One respondent wrote ``[I love that /r/blackladies] is a community of black women coming together to discuss social issues that are prevalent and important to us.'' Another says ``[/r/Glaucoma] lets me get in touch with people around the world [who are] dealing with a similar health issue.'' These quotes from participants reflect a body of literature that finds that online communities can be helpful venues for minority groups and those with special needs connect with similar people for support~\cite{Raj2017PsychosocialSF, Kamalpour2020HowCO}.

\subsection{Size}\label{val:size}
Many responses (144/1481) commented on the \textit{Size} of communities, both regarding the \textit{Volume of Content} submitted to the community and the number of people in that community (\textit{Size of Community}). 
\new{Participants were varied in their preferences, with some preferring a larger volume of content}
(\eg ``I like that /r/assholedesign is regularly updated'')
\new{and others preferring less content.}
One member of /r/SampleSize, a community for distributing surveys, wrote ``The large volume of posts means that you need to time your submission very carefully in order to make sure that your survey doesn't get buried.''

\new{Similarly,} respondents were fairly evenly split (21 preferring larger, 15 preferring smaller) on their desire for larger or smaller communities. Those preferring larger communities perceived those larger communities as offering increased opportunities for interaction (``More active users [on /r/photocritique] would make it easier to engage with like minded individuals through their posts.''), while those preferring smaller communities tended to like the specificity of \new{subject matter} and focus, \eg ``[I like that /r/roguelikedev, a community dedicated to the development of a specific type of video game, is a] tiny community with a very clear purpose and scope.''
\new{Community members' value \new{preferences} regarding community size seem likely to depend on many factors, especially including the subject of the community and the nature of interactions which occur within.}

\subsection{Participation \& Inclusion}\label{val:inclusion}
\textit{Participation and Inclusion} in the community was a value \new{topic} reported by 93 respondents. We divide their responses into three subcategories: \textit{Tools for Participation}, \textit{Offensive/Abusive/Harassing Content or Behaviors}, and \textit{Outsiders/Demographics/Limits}, which focuses on \textit{who} participates.

Comments on \textit{Offensive/Abusive/Harassing Content or Behaviors} were split (20 units to 38 units) between positive comments (praising the absence of offensive content), \eg ``[I like that /r/WANDAVISION] has no homophobia, racism, or any discrimination,'' and comments that described the respondent's experience with such behavior that was detrimental to inclusion. One member of /r/sewing wrote ``sometimes the commenters on a post will write personal things about a poster's appearance, which I don't think is appropriate."

Community members who commented on \textit{who} participates in their communities (\textit{Outsiders/Demo\-graphics/Limits}) were frequently concerned with the presence of outsiders, who are perceived as not belonging to the community by virtue of their lack of familiarity with the \new{subject} or even their physical location. One member of /r/Seattle wrote that ``too many people from outside Seattle comment on posts in this subreddit.'' 
In any community with a specific \new{subject}, some degree of boundaries for membership are natural, yet too much insularity in online communities can be harmful~\cite{Allison2020CommunalQA}.

\textit{Tools for Participation} were suggested by seven participants who felt their communities were lacking such tools. One participant wrote ``[/r/snails] should add an easily accessible ‘beginners’ questions’ section.''
Some tools, such as automated posts and messages~\cite{Yazdanian2019ElicitingNW} and badges, have been found to reduce unwelcoming reactions to new community members~\cite{santos_2020_badges} as well as such members' compliance with rules~\cite{matias_preventing_2019}.

\subsection{Diversity}\label{val:diversity}
Many community members commented on the \textit{Diversity} of their communities, which we divide into two subcategories: \textit{Variety of Content} and \textit{Diversity of People}. More respondents (56/75) commented on \textit{Variety of Content}, \eg ``[/r/CollegeBasketball]  has an ideal blend of banter, rumors, statistical insights, and glorious shitposting.'' Those who commented on \textit{Diversity of People} more frequently commented on aspects of the community they would like to change, \eg ``I wish [/r/knitting] had more variety in skill. Right now it's mostly skilled knitters, whereas it would be appreciated to see some beginner knitters.''

\subsection{Moderation \& Moderators}\label{val:mods}
\textit{Moderation \& Moderators} are controversial \new{subjects} on reddit, and many moderators feel strongly disliked by members of the community they moderate~\cite{matias_civic_2019}. However, close to half (31/69) of idea units regarding moderation were positive, with respondents praising the moderator team in general (``mods [of /r/Phillipines] have done an excellent job of maintaining the community'') as well as specific moderators (``$<$username redacted$>$ is such a great mod [of /r/ApplyingToCollege] who is super helpful.'').

Of the 38 negative comments on moderation, many of them were critical of perceived arbitrary rule enforcement, \eg ``moderators arbitrarily remove posts because they're `against the subreddit's rules'.'' 
Other idea units requested more active moderation (14 units), complained about perceived power abuses (8 units), and called for greater community involvement in rule making (2 units).

\subsection{Norms, Voting Behavior, and Adherence}\label{val:norms}
Almost every (54/59) response regarding \textit{Norms} was on a negative aspect of the community, suggesting that norms are mostly noticeable in a community when they are violated.
Many of the idea units relating to norms (46/59) were community specific and focused on \textit{Adherence to Norms}, such as newcomers not reading the FAQ before posting, posters not providing adequate information and expecting a response (/r/whatisthisplant), or not including appropriate sources. Many respondents also requested additional rules or changes to norms (19 units), \eg ``I would like to reduce the amount of people speaking English [on /r/ich\_iel, a German-speaking subreddit].''

The remainder of idea units (10/59) were complaints about \textit{Voting Behavior}, particularly the use of the ``downvote button as a disagree button.'' 

\subsection{Technical Features}\label{val:technical_features}
53 idea units (3.6\%) focused on \textit{Technical Features}, with three subcategories: \textit{Flairs/Tags/NSFW\footnote{A common reddit acronym meaning `not safe for work' and used to indicate content containing gore or nudity.} Labels}, \textit{Search and Filters}, and \textit{Recommendation Systems}. Most (36/53) idea units in \textit{Technical Features} focused on negative aspects of technical features, particularly such features' absence or failure to work properly. Most positive idea units praised their respective subreddits' use of flair and tags, \eg ``NSFW tagging is perfectly used here [in /r/hemorrhoid].''

While flairs, tags, recommendations systems, filtering options, and quality search functionality can dramatically improve users' experiences in online communities, on reddit, such technical features can be difficult for communities to implement and modify, as they are controlled by the reddit administration, not community members. Frequently, incentives are misaligned between subreddit leadership and the reddit administration, who may be unwilling to implement new technical features at the request of communities. As a workaround, many communities implement their own second-party technical features by modifying or repurposing existing features, such as custom community moderation bots~\cite{Kiene2020WhoUB} or reputation systems such those in /r/changemyview~\cite{Jhaver2017DesigningFC}.

\subsection{Trust}\label{val:trust}
The \textit{Trust} of a community was the least frequently commented on \new{value category}. This category consists of two subcategories, \textit{Knowledgeable People} and \textit{Trustworthy Content}, again differentiated by a focus on people \textit{vs.} content. Of respondents who commented on the trustworthiness of people, one respondent appreciated the credentials of community members ``There are many doctors [on /r/eyetriage] so it's a lot more reliable,'' whereas another respondent lamented how anonymity interfered with the trustworthiness of the /r/MachineLearning community: ``Anonymity sometimes makes it so I don't know who is qualified to say what.''

\section{Implications \& Discussion}\label{sec:discussion}
Community members value a broad range of \new{topics} for their communities ranging from \textit{Diversity} (\sect\ref{val:diversity}) to \textit{Technical Features} (\sect\ref{val:technical_features}).
However, some of these \new{topics}, such as the \textit{Quality of Content} (\sect\ref{val:quality}), are much more frequently reported by community members than other\new{s}.
We do not, however, suggest that these \new{topics which are most frequently mentioned} should be considered more important to \textit{researchers} than less frequently \new{mentioned topics}, such as \textit{Trust} (\sect\ref{val:trust}).
Most existing research that seeks to make online communities `better' focuses on implicit community values which are relatively narrow in scope, especially those with broader societal impact (\eg mis/disinformation and political polarization) and/or a negative impact on vulnerable populations (\eg abuse, harassment) (\sect\ref{sec:existing_comparison}).
While these research directions are critical to mitigating major harms associated with online communities, our findings suggest that many values held by community members are understudied, complex, and often in conflict with one another.
In this section we discuss this complexity (\sect\ref{sec:understudied_values}) and conflict (\sect\ref{sec:conflicting_values}),  implications for moderation and governance practices (\sect\ref{sec:governance_discussion}), and \new{subsequently} how our taxonomy relates to other taxonomies from different contexts (\sect\ref{sec:existing_comparison}).

\subsection{Understudied Community Values}\label{sec:understudied_values}
Many values explicitly stated by community members \new{have topics which are} are not well studied by the research community. We find that \textit{Quality of Content} is the most frequently mentioned \new{value topic} (\sect\ref{val:size}), accounting for 46.5\% of idea units in our responses, yet with the exception of spam and bot detection, this value is poorly understood. Quality of content is arguably the most challenging value \new{topic} to define and measure, as it is extraordinarily context-specific and dependent on the nature of the community in question. \new{For example,} content that is high quality in /r/catpictures would be woefully inappropriate for /r/science. As existing work in this space focuses mostly on understanding the quality of conversational (and other text-based) content~\cite{Zhang2018CharacterizingOP, Zhang2018ConversationsGA, lakkaraju_whats_nodate}, additional work which contributes to the understanding of the context-specific quality of image and video-based content is especially needed.

Community \textit{Size} is another value \new{category} that is understudied and complex. Some work has studied the impact of rapid growth on communities~\cite{lin_2017_better_when_smaller, kiene_2016_eternal_september}; however, our findings suggest that members perceive a difference between changes in the size of the community (\eg the number of participants) and the volume of content (\sect\ref{val:size}). While most respondents (47/76) \new{prefer} a larger volume of content, many respondents (15/36) perceive smaller communities as better due to having stronger community engagement and interactions, \eg ``as the community has grown [it] has lost its small and friendly community feel''. Balancing this tension is a important \new{area} for future research, as is understanding how desired community size varies as a function of the community member's relationship with the community in question.

\subsection{Conflicting Community Values}\label{sec:conflicting_values}
\new{
Our results show that community members hold a broad range of values, but these values are challenging to implement because values in different categories often conflict with one another. 
}

\xhdr{Inclusion vs. Quality of Content and Norms}
While the challenge of incorporating new members has been previously identified and studied~\cite{robert_e_kraut_building_2012, dnm_2013_old_members, cho_2021_potential_new_members}, our findings permit the framing of this tension as a conflict between \textit{Inclusion} (of new members) and \textit{Quality of Content} and \textit{Norms}.
One participant expressed this sentiment in their response, saying ``It's frustrating when [new members] tend to leave out information they're expected to include.''
In social media communities, some work has experimented with onboarding documents and mentorship~\cite{santos_2020_badges, matias_preventing_2019, Yazdanian2019ElicitingNW} to improve new members' understanding of community norms and maintain the quality of content. Methods to mitigate the tension between inclusion, norms, and quality of content have been studied more deeply in the context of peer production communities such as Wikipedia~\cite{ciampaglia_2015_moodbar, halfaker_rise_2013, halfaker_2011_bite_newbies}, and there is great potential for future work to study how these findings generalize to more social communities and develop new tools~\cite[Ch.5]{robert_e_kraut_building_2012}. To an extent, however, these values are inherently at odds.

\xhdr{Size vs. Community Engagement}
Community engagement and size are also often in conflict with one another. While several studies have found that, by some metrics, communities' health is not harmed in the long term by increases in size~\cite{lin_2017_better_when_smaller, kiene_2016_eternal_september}, growth is a frequent subject of complaint across many platforms. One of our participants wrote ``[/r/formula1 is] such a large community that is hard to engage with other members.''
On the other hand, many participants also reported desiring more activity and more frequent content in their communities. An open and important research question is how to scale communities to larger size without sacrificing the sense of interpersonal engagement.

\subsection{Community Governance \& (Lack of) Consensus}\label{sec:governance_discussion}
\new{
While the previous section described conflicts between different value categories, we also find evidence that members of the same community can disagree with one another's value \textit{preferences} for the same value category.
\new{For example,} eight participants who are members of /r/AskReddit expressed differing preferences for what content is permitted in the community. Three idea units indicated a preference for an `anything goes' strategy, while the remaining four wished for more restrictions on specific \new{subjects}, mainly sex and drug use.
}
Other research has found similar evidence that community members often disagree with one another on \new{matters} such as the severity of harmful behavior~\cite{scheuerman_2021_framework_severity, jiang_2021_international_harm}.
Strategies for reconciling these differences of opinion, however, are not well understood.
In the context of peer production communities such as Wikipedia, some research has already explored some sources of internal disagreement, such as tensions between senior and junior members~\cite{halfaker_2011_bite_newbies, halfaker_rise_2013, steinmacher_2015_barriers_newcomers}, however the extent to which these findings generalize to social media platforms such as reddit may be limited.
Our results suggest that even relatively small samples of members are adequate to surface some differences in values between community members.
Future work should examine the degree of agreement or disagreement on values held by many members from the same communities, and what factors are predictive of such differences in opinion.

\xhdr{Affordances for Participatory Governance}
Participatory governance practices can help build consensus and reconcile differences in opinion when considering and implementing rule changes. However, affordances for such practices are extremely limited on social media platforms, where the vast majority of communities' rules are determined exclusively by a small set of moderators with no formal input from the broader community~\cite{zhang_policykit_2020}.
We find (\sect\ref{val:mods}) that many community members perceive lack of transparency and arbitrary enforcement decisions as evidence of corruption (``mods will ban you without warning if you say something they disagree with''), and desire greater input into moderation (``moderators should consult the community about what we want'').
Conflict between moderators and the broader community has been identified in some prior work~\cite{matias_civic_2019}, yet the exact differences between moderators' and nonmoderators' opinions have not yet been studied and quantified systematically. Increasing participatory governance practices in online communities may help alleviate some of this conflict, yet such practices are not a panacea, and can cause harm if not implemented carefully, \eg by increasing the burden of labor on certain groups, or by giving a veneer of legitimacy to unilateral decision making~\cite{Kelty2017TooMD}.

\xhdr{Methods for Managing Irreconcilable Differences}
What happens when differences in values are too great to reconcile? On reddit, it is not uncommon for some members to splinter and create an alternative subreddit in response to perceived grievances or frustration with the status quo, however, to the best of our knowledge, this `exit' phenomenon has not been studied in depth~\cite{frey2021exit}. Other communities, such as consensus-based peer production communities, have different practices to manage internal disputes such as formal arbitration committees~\cite{Konieczny_2017_wiki_arbitration}, but it is unclear how such practices would generalize to social media communities. In some cases of divergent values, such as differing \textit{Personal Preferences} for content, personalized filtering may be used so that each user does not see content they wish to avoid~\cite{jhaver_2018_blocklists}. However, this undermines social translucence, a theory which suggests that making online behavior visible creates social spaces with shared accountability~\cite{Erickson_2000_social_translucence}. Some research has explored interventions to balance the trade-off between social translucence and the personalization afforded by filtering~\cite{gilbert_2012_social_translucence}.

\xhdr{Power Structures and Protecting Vulnerable Community Members}
Furthermore, it is likely that some of the \new{categories of} values we have identified in this work will be of special importance to vulnerable groups. For example, one member of /r/AskWomenOver30 said ``everyone assumes I'm a white American, which really changes the dynamic when I ask about career, relationships, and more.''  In this case, only a small fraction of the community may recognize that these assumptions are harmful to the experience of some community members, but that minority perspective is still very important.
If traditional participatory decision making practices such as voting were implemented naively, the majority (which may not have had personal experience with harms such as the aforementioned assumptions of members' background, harassment~\cite{matias_study_2020}, \textit{etc.}) may not support or even be opposed to measures designed to reduce these harms, which disproportionately affect women and other minority groups~\cite{Lenhart2016OnlineHD}.

This phenomenon, known as `Tyranny of the Majority,' is well known in the offline governance context~\cite{Guinier1994TheTO}, yet has not been studied in-depth in the online context. Some peer production communities, such as Wikipedia, emphasize consensus-based deliberation over vote-counting partly as a way to avoid this issue~\cite{wikipedia_polling_is_not_a_substitute_for_discussion, im_2018_wikipedia_rfc}, though issues with bias against women and other minority groups still persist~\cite{tripodi_2021_ms_categorized}.
Additional research into participatory governance practices for online communities which protect vulnerable groups is sorely needed.

At a higher level, much research on online communities studies these communities through the lens of moderators~\cite{matias_civic_2019} and other people in positions of power~\cite{robert_e_kraut_building_2012, chandrasekharan_2017_you, shen_discourse_2019, habib_act_2019}. This is especially the case for empirical work which relies upon rules and enforcement actions as concrete evidence of behavior and community norms~\cite{jhaver_2019_user_reactions, jhaver_does_2019, chandrasekharan_internets_2018, fiesler_reddit_nodate}. While these rules and enforcement actions are natural sources of empirical data, researchers must take care and recognize that many community members feel as though they do not have an adequate stake in rulemaking and enforcement (\sect\ref{val:mods}). Online platforms' power structures are complex~\cite{jhaver_2021_designing}, and while surveys such as this one are useful tools, additional work is needed on scalable methods for empirically studying community behavior and values that include the voices of the least empowered.

\subsection{Limitations}\label{sec:limitations}
While we made intentional efforts to recruit participants from a diverse set of backgrounds by using multiple recruiting methods, it is possible that selection effects impacted who responded to our survey, which may have resulted in a taxonomy that is not truly representative of all community members.
Furthermore, we do not include participants younger than the age of 18.
Lastly, while our taxonomy derived from responses relating to a large set of 627 diverse communities, we only have responses from a small fraction of each community's membership.
These facts, along with our (in absolute terms) low click-through rate of 0.227\%, are indicative of a fundamental reality for researchers of online communities: it is very difficult to effectively poll an online community. Additional research is needed to improve polling methods, \eg by reducing friction in online surveys.
\new{Improved methods could strengthen} results' representativeness and susceptibility to bias from sources such as poorly defined community membership.

We believe the validity of our taxonomy is demonstrated both by the large (1,180 idea unit) validation set used to demonstrate saturation (\sect\ref{sec:categorization}), the high inter-rater reliability (Fleiss kappa = 0.874) of the codebook, as well as the correspondence between the values \new{reported by our participants} and those identified by sociologists and psychiatrists in the context of interpersonal interactions (\sect\ref{sec:existing_comparison}). Making `truly representative' samples of online communities' membership is especially difficult given that such membership is often not well-defined, yet this is an important \new{area} for future work.

\begin{table*}[ht]
\centering
\small
\begin{tabularx}{\linewidth}{X|X|X|X}
\textbf{This Work}                                                   & \textbf{\citet{bao_conversations_2021}}                                   & \textbf{\citet{deri_coloring_2018} and \citet{choi_ten_2020}}             & \textbf{\citet{fiesler_reddit_nodate}}                                                                                                 \\
Iterative categorization of 212 redditors' survey responses & Survey of Prosocial Behavior literature                                   & Survey of Sociology \& Psychology literature & Manual coding of 300 rules from 18 subreddits                                                                    \\ \hline\hline
Quality of Content                                                   & Information Sharing                                                       & Knowledge, Fun                               & Content/Behavior, Format, Low-Quality Content, Off-topic, Reposting, Spam, Advertising/Commercialization, Images \\ \hline
Community Engagement                                                 & Social Cohesion, Social Support, Gratitude, Mentoring, Esteem Enhancement & Power, Status, Support                       & Personality                                                                                                      \\ \hline
Diversity                                                            & Social Cohesion                                                           & Similarity, Identity                         & Off-topic                                                                                                        \\ \hline
Size                                                                 & Social Cohesion                                                           & N/A                                          & N/A                                                                                                              \\ \hline
Participation \& Inclusion                                           & Social Support, Mentoring, Absence of Antisocial Behavior                & Trust, Identity, Conflict                     & Doxxing/Personal Info, Harassment, Hate Speech, Trolling, Links \& Outside Content                               \\ \hline
Technical Features                                                   & N/A                                                                       & N/A                                          & NSFW                                                                                                             \\ \hline
Moderation \& Moderators                                             & N/A                                                                       & Power, Status                                & Consequences/Moderation/ Enforcement                                                                             \\ \hline
Norms                                                                & Absence of Antisocial Behavior                                            & Trust, Conflict                              & Format, Voting                                                                                                   \\ \hline
Trust                                                                & Information Sharing                                                       & Trust                                        & Content/Behavior                                                                                                 \\ \hline
Not mentioned by community members                                   & Fundraising \& Donating                                                   & Romance                                      & Copyright/Piracy, Personal Army, Politics, Sitewide                                                                 
\end{tabularx}
\caption{A comparison of how categories from our empirically-derived taxonomy are mapped onto by taxonomies from prior work on small group interactions~\cite{bao_conversations_2021, deri_coloring_2018, choi_ten_2020} and community rules~\cite{fiesler_reddit_nodate}. All of our categories have analogues in these other taxonomies from different contexts.}
\label{tab:comparison_of_dimensions}
\end{table*}

Our participants were asked about their experiences on a single social platform, reddit. While reddit is large, popular, and contains many thousands of communities covering diverse \new{subjects}~\cite{medvedev_anatomy_2019}, reddit has important differences from other platforms. Unlike reddit, platforms such as Twitter and Facebook lacks explicit communities with well-defined membership.
reddit has a stronger focus on link-sharing than other platforms, in part due to technical differences and in part due to history and the culture on the platform.
reddit is also almost entirely public, unlike private communities on Slack, Discord and some Facebook Groups.

Reviews of research conducted using reddit data have found that the generalizability of results from reddit to other contexts are promising, with certain caveats, and this is an important area for additional research~\cite{proferes_2021_studying_reddit}. However, we believe that the size and prominence of reddit means that our work still has an important impact, regardless of generalizability.

reddit also differs dramatically from peer production communities such as Wikipedia and open source projects, where the community has a clear focus beyond social engagement and entertainment. Future work is needed to understand how community values may differ in these other contexts.

reddit also mostly consists of English-speaking users from Western cultures, and additional work is needed to understand how people from other cultures may hold different values. Some evidence for cultural differences in values has already been reported~\cite{jiang_2021_international_harm}.

\section{Related Work \& Comparison to Existing Taxonomies}\label{sec:existing_comparison}

\xhdr{Community Rules, Norms, \& Content Moderation}
One way in which \new{commonly-held} community values manifest is in the formalized rules of communities and how those rules are enforced by content moderators. Rules on reddit have been examined in prior work \cite{fiesler_reddit_nodate, chandrasekharan_internets_2018}, however, as subreddits' rules are written and enforced exclusively by community moderators, they may not be representative of the values of the broader community membership~\cite{matias_civic_2019}.
\new{
On reddit, as on most social media platforms, more democratic self-moderation is relatively rare~\cite{seering_2020_self-moderation}, with platforms' technical features built mostly around a strictly defined hierarchy of admins and moderators, encouraging an `implicit feudalism'~\cite{Schneider2021AdminsMA}. In these cases, conflict between non-moderator membership and the moderators over rules and their enforcement is relatively common~\cite{matias_civic_2019, jhaver_2019_user_reactions, srinivasan_2019_content_removal}. 
While some third party tools to enable a broader range of governance practices have been proposed~\cite{zhang_policykit_2020}, these tools are not yet widely adopted.
}
In this work, we directly ask all community members about their values, not just moderators.


\xhdr{Intracommunity Tension and Conflict}
\new{
Conflict between non-moderators and moderators is not the only form of tension within online communities. One major challenge in communities is that of integrating new members into an existing community~\cite[Ch.5]{robert_e_kraut_building_2012}. This has been studied empirically on a wide range of platforms, finding that new members mostly learn community norms from experience~\cite{cho_2021_potential_new_members}, and that once established in a community, members are less likely to change their habits~\cite{dnm_2013_old_members}.
As a result, periods of massive growth can result in substantial change to the community~\cite{lin_2017_better_when_smaller, halfaker_rise_2013} and frustration amongst existing members~\cite{kiene_2016_eternal_september}.
In communities that are especially focused on topics requiring special knowledge or expertise, a related tension often occurs between those with greater knowledge and those without; this has been studied on reddit in case studies for science~\cite{jones_2019_rscience} and history communities~\cite{gilbert_2020_askhistorians}.
In cases of extreme intracommunity conflict, such tensions can even lead to fragmentation as some members exit~\cite{frey2021exit} and form new, alternative communities~\cite{Fiesler2020MovingAL, Newell2016UserMI}.
Additional evidence for intracommunity tension can be found in work that examines how members define rulebreaking and how to fairly punish such behavior; this work has found substantial disagreement amongst community members~\cite{scheuerman_2021_framework_severity, jiang_2021_international_harm}.
}
\new{
Our taxonomy enables us to frame these tensions as conflicts between different \new{categories of} values (\sect\ref{sec:conflicting_values}), \new{or as differences in members' value preferences in the same value category}.
}

\xhdr{Implicit Values in Prior Research}
Any research that seeks to improve the health of an online community implicitly (and often explicitly) values certain aspects of that community.
For example, the abundance of research on reducing misinformation \cite{bovet_influence_2019, weld_political_2021, grinberg_fake_2019, anagnostopoulos_viral_2014} implicitly values the veracity of content.
Similarly, research seeking to reduce harassment \cite{burke_winkelman_exploring_2015, matias_study_2020} implicitly values safety.
This `implicit values' perspective can be used to identify what values of online communities are most studied by the research community, and conversely, what values are most understudied (\sect\ref{sec:understudied_values}).


In this work, we do not argue that the set of values derived from community members' responses is superior to any other set of values, but instead conduct a survey to outline the diversity of values held by community members, how they may be in conflict with one another, and how they relate to prior research and can inform future research.

\xhdr{Comparison to Existing Taxonomies}
While we are not aware of any work prior to ours that directly asks members of online communities their values, aspects of online interactions have been studied in the context of 1:1 or small group interactions (\citet[Table 1]{deri_coloring_2018} provides an overview). 
We summarize how the major categories from our results (\sect\ref{sec:taxonomy}) relate to categories in taxonomies of aspects of small group interactions~\cite{bao_conversations_2021, deri_coloring_2018, choi_ten_2020} and the taxonomy of subreddit rules from \citet{fiesler_reddit_nodate} in Table~\ref{tab:comparison_of_dimensions}.

Every major category in our taxonomy has at least one analogue in the small group interaction and rule taxonomies. Naturally, given the different contexts, there are differences in how and where the categories overlap, and some components of the other taxonomies are not as relevant to the large online community context we study here.
The small group taxonomies emphasize dynamics that affect 1:1 interactions, such as gratitude, mentorship, and asymmetrical power dynamics. In large communities, these translate to the quality of community engagement and interaction, the inclusion of new members, and the power wielded by moderators, respectively. Romance and Fundraising/Donating, which appear in the small group taxonomies, do not appear in our taxonomy and were not mentioned by any of our 1,481 participants' idea units.

The taxonomy of different rules on reddit~\cite{fiesler_reddit_nodate} also overlaps substantially with our taxonomy, as rules are a formalized reflection of community values. As rules are primarily written with regards to content~\cite{fiesler_reddit_nodate}, our \textit{Quality of Content} category is relevant to many categories from the rules taxonomy. Categories from the rules taxonomy which do not have analogues in our taxonomy are rules regarding copyright/piracy, politics, and `personal armies' (brigading, when large numbers of members from one community temporarily participate in an often hostile manner in another community).
\section{Conclusion}\label{sec:conclusion}
Online social communities are rich spaces that can bring people together in a healthy, productive, and enjoyable manner. Many researchers study how to make online communities `better,' but understanding what `better' means is a challenging problem, as there is no single set of values for online communities that can be used to inform research in this space. The values held by community members themselves are difficult to measure, and their perspectives have mostly not been included in existing research.

In this work, we surveyed 212 redditors who are members of 627 unique communities. Using open ended questions (\sect\ref{sec:instrument}), we asked these redditors what their values for their communities are, in their own words. Using an iterative categorization method based in grounded theory (\sect\ref{sec:categorization}), we contributed a taxonomy of 29 \new{subcategories of community values} across a broad range of topics from the diversity of the community to technical features (\sect\ref{sec:taxonomy}).
Raters using our codebook demonstrate very high inter-rater agreement (Fleiss kappa = 0.874).

Our findings have important implications for future work on online social communities, and have already enabled followup work~\cite{weld_2022_survey_icwsm, Maftouni2022ThankYF}. We highlighted understudied and challenging-to-implement community values such as \textit{Quality of Content}, \textit{Size}, and \textit{Community Engagement} (\sect\ref{sec:understudied_values}). We identified where community values conflict with one another (\sect\ref{sec:conflicting_values}), and called for additional work on participatory governance for online communities that protects vulnerable groups of community members (\sect\ref{sec:governance_discussion}).

\section*{Acknowledgements}
This research was supported by the Office for Naval Research (\#N00014-21-1-2154). NSF grant IIS-1901386, NSF CAREER IIS-2142794, NSF grant CNS-2025022, the Bill \& Melinda Gates Foundation (INV-004841), and a Microsoft AI for Accessibility grant.

{
\small
\bibliography{bib_trimmed}
}

\clearpage
\onecolumn
\appendix
\section{Additional Figures Describing Survey Interface and Recruiting}\label{app:extra_figs}

\begin{figure}[!ht] 
    \centering
    \includegraphics[width=.7\textwidth]{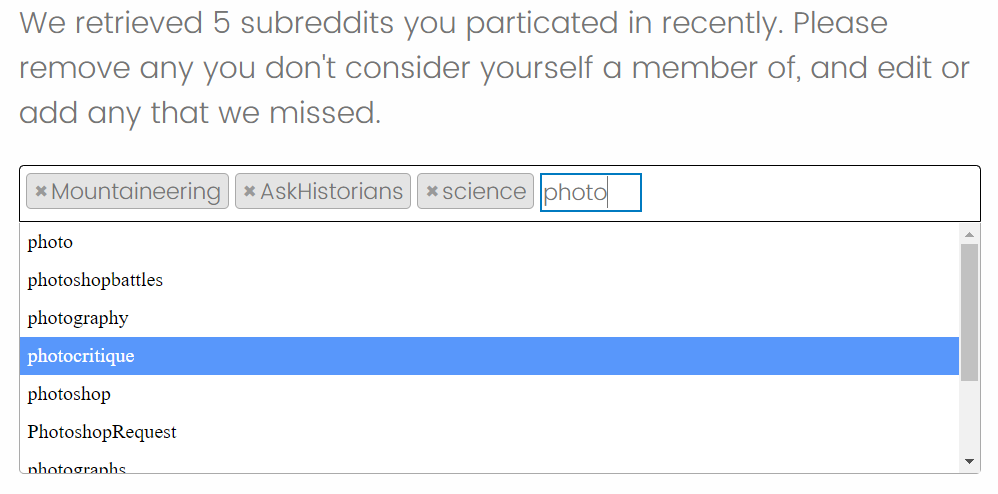}
    \caption{A screenshot of the interface used by participants to enter the subreddits they consider themselves a member of. This search box queries the reddit API in real-time to populate the results and ensure that only valid subreddit names are entered.}
    \label{fig:subreddit_selector}
\end{figure}

\begin{figure*}[h]
     \centering
     \begin{subfigure}[b]{0.45\textwidth}
         \centering
         \includegraphics[width=\textwidth]{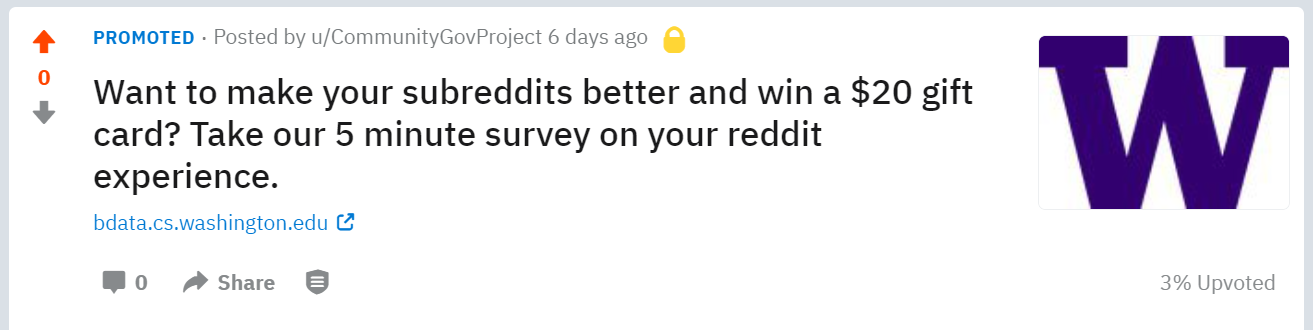}
         \label{fig:ad_1}
     \end{subfigure}
     \hfill
     \begin{subfigure}[b]{0.45\textwidth}
         \centering
         \includegraphics[width=\textwidth]{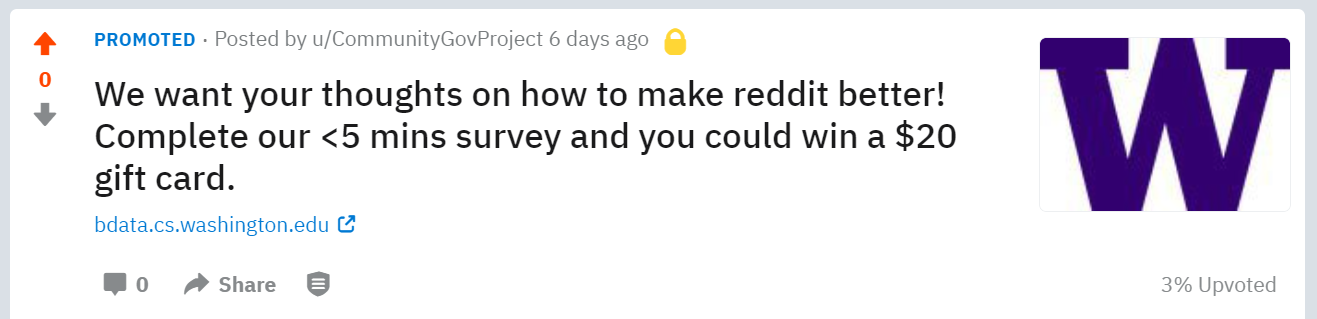}
         \label{fig:ad_2}
     \end{subfigure}
     \caption{Reddit advertisements used to recruit participants.}
     \label{fig:recruitment}
\end{figure*}

\newpage

\section{Additional Recruiting and Incentive Details}\label{app:recruiting}
Participants for this study were recruited primarily through the purchase of reddit advertisements. These advertisements display inline with other content in both individual subreddits and aggregated content views, and are shown on the reddit website as well as the official reddit mobile app.
Figure~\ref{fig:recruitment} shows the appearance of these advertisements.

Upon clicking on the advertisement, the user is taken to the first page of the survey which summarizes the aims of the study and requests informed consent to continue.
Over the course of the study, 920,025 advertisement impressions were made, generating 2,084 clicks, for a click-through rate of 0.227\%. Of these people who clicked through to the first page of the survey, 509 started to complete the survey, and 212 completed it. Eleven respondents did not consent to continuing the survey, and as a result were not shown any additional questions.
To increase the diversity of recruiting, the survey was also distributed to relevant university mailing lists and Slack channels at two large American universities, as well as posted to /r/SampleSize, a subreddit for the distribution of surveys, recruiting an additional 41 participants.

To incentivize respondents to participate, a raffle was held, and winners were chosen at random from the pool of respondents who both completed the survey and supplied their reddit username, which was used to contact the winners. One `first place' prize, a \$100 Amazon gift card, and five `second place' prizes of \$20 Amazon gift cards were awarded (equivalent amounts in local currency were provided to participants outside of the United States).

\section{Additional Details on Participant Demographics}\label{app:demographics}

\begin{table}[!ht] 
\small
\centering
\begin{tabular}{lrrr}
\hline
                                         & \multicolumn{2}{r}{\textbf{Our Survey}} & \textbf{reddit Overall}              \\
                                         & \multicolumn{2}{r}{($n = 212$)}         & \citet{pew_2016_reddit_demographics} \\ \hline
\textbf{Age}                             &                                         &                                      \\
18-29                                    & 101   & (71.1\%)                        & 59\%                                 \\
30-49                                    & 35    & (24.6\%)                        & 33\%                                 \\
50-64                                    & 6     & (4.2\%)                         & 7\%                                  \\
65 or older                              & 0     & (0.0\%)                         & \textless{}1\%                       \\ \hline
\textbf{Gender}                          &       &                                 &                                      \\
Woman                                    & 49    & (26.2\%)                        & 33\%                                 \\
Man                                      & 129   & (69.0\%)                        & 67\%                                 \\
Non-binary                               & 9     & (4.8\%)                         & \multirow{2}{*}{Not reported}        \\
Additional Gender                        & 5     & (2.7\%)                         &                                      \\ \hline
\textbf{Race and Ethnicity}              &       &                                 &                                      \\
White (Hispanic, Latino, or Spanish)     & 32    & (16.9\%)                        & \multirow{2}{*}{7\% Hispanic}        \\
Non-white Hispanic, Latino, or Spanish   & 6     & (3.2\%)                         &                                      \\
White (Not Hispanic, Latino, or Spanish) & 89    & (47.1\%)                        & 74\% White                           \\
Black or African American                & 3     & (1.6\%)                         & \multirow{2}{*}{8\% Black}           \\
Middle Eastern or North African          & 8     & (4.2\%)                         &                                      \\
Asian                                    & 46    & (24.3\%)                        & \multirow{4}{*}{10\% Other}          \\
Native Hawaiian or Pacific Islander      & 3     & (1.6\%)                         &                                      \\
Indigenous American or Alaskan Native    & 1     & (0.5\%)                         &                                      \\
Additional Race/Ethnicity                & 10    & (5.3\%)                         &                                      \\ \hline
\end{tabular}
\caption{Demographics of survey respondents and overall reddit demographics. In general, our respondents' demographics are similar to the overall demographics of reddit. Participants could choose multiple gender and race/ethnicity options, so percentages may not sum to 100\%. Overall reddit demographic data \cite{pew_2016_reddit_demographics} uses different racial identity questions, so while an exact comparison is not possible,  similar categories are provided here.}
\label{tab:demographics}
\end{table}

\newpage
\section{Survey Instrument}\label{app:survey}












{
\setlength{\parindent}{0cm}

\subsection*{Informed Consent}

This survey aims to learn more about the subreddits that reddit users participate in, and what values redditors have for those subreddits.

In this survey, we will ask you a few questions about your overall reddit usage, and then ask you a few questions each about the subreddits you consider yourself to be a member of. You may skip any questions you'd prefer not to answer. It should take less than 5 minutes to complete.

Only high-level data will be published as part of our research. Your responses are confidential, and will never be made public.

This study is run by researchers from the University of Washington, 
and has been determined to be exempt from IRB approval under University of Washington IRB ID STUDY00011457.
For questions or concerns, please contact gweld@cs.washington.edu, or /u/cyclistNerd on reddit.

Would you like to participate in this survey?

\begin{itemize}
    \item[$\circ$] Yes, I would like to participate in this survey.
    \item[$\circ$] No, I would not like to participate in this survey.
\end{itemize}

\subsection*{Reddit Username and Compensation}
As compensation for your participation in this study, you will be entered in a raffle for one of five Amazon gift cards, worth
\$20 each.

To contact you after the raffle drawing, we will send you a reddit private message.
To do so, we ask for your reddit username.

We will also use your username to identify your posts and comments in the subreddits you participate in.
Your username will be kept confidential, and we will never publish any of your reddit history.

Providing your username is entirely optional, but without it, we cannot enter you into the raffle.

What is your primary reddit username? Your answer will be kept confidential.

Please spell carefully, and do not include `/u/' or `u/'.

\fbox{\phantom{This is my username.}}

Do you have multiple reddit accounts?

\begin{itemize}
    \item[$\circ$] No, I only have one reddit account.
    \item[$\circ$] Yes, I have multiple reddit accounts.
\end{itemize}

You entered that your primary username is /u/\underline{\hspace{2cm}}

If this looks correct, press next to start the rest of the survey.
If this is incorrect, please press back to edit your response.

\subsection*{Multiple Accounts}

Earlier, you indicated that you had multiple reddit accounts.

If you're comfortable doing so, please enter the names of your alternate reddit accounts, separated by commas.

\fbox{\phantom{These are my other usernames.}}

\subsection*{Demographics}

What is your age (in years)? \fbox{\phantom{xx years old.}}

Please describe your gender (check all that apply)
\begin{itemize}
    \item[$\square$] Woman
    \item[$\square$] Man
    \item[$\square$] Non-binary
    \item[$\square$] \fbox{\phantom{I identify as }} Prefer to self-describe
\end{itemize}

Please describe your race (check all that apply)
\begin{itemize}
    \item[$\square$] White (Hispanic, Latino, or Spanish)
    \item[$\square$] White (Not Hispanic, Latino, or Spanish)
    \item[$\square$] Non-white Hispanic, Latino, or Spanish
    \item[$\square$] Black or African American
    \item[$\square$] Asian
    \item[$\square$] Middle Eastern or North African
    \item[$\square$] Native Hawaiian or Pacific Islander
    \item[$\square$] Indigenous American or Alaskan Native
    \item[$\square$] \fbox{\phantom{I identify as }} Prefer to self-describe
\end{itemize}

\subsection*{Overall Reddit Usage}
In this section, we will ask you about how you use reddit.

Typically, how often do you use reddit?
\begin{itemize}
    \item[$\circ$] Every day.
    \item[$\circ$] A few times a week.
    \item[$\circ$] Once a week.
\end{itemize}

Typically, how long do you typically spend on reddit at one time?
\begin{itemize}
    \item[$\circ$] Less than five minutes.
    \item[$\circ$] More than five minutes, but less than fifteen minutes.
    \item[$\circ$] More than fifteen minutes, but less than an hour.
    \item[$\circ$] More than an hour.
\end{itemize}

Typically, how often do you post or comment on reddit versus browsing what others have submitted (lurking)?
\begin{itemize}
    \item[$\circ$] Frequently, I frequently submit posts or comment on threads.
    \item[$\circ$] Occasionally, I post or comment occasionally, but mostly browse what others have submitted.
    \item[$\circ$] Rarely, I almost always just browse what others have submitted.
    \item[$\circ$] Never, I only browse what others have submitted.
\end{itemize}

How often do you use aggregate subreddits (like your frontpage, /r/all, or multi-reddits) versus looking at individual subreddits?
\begin{itemize}
    \item[$\circ$] Always, I never look at individual subreddits.
    \item[$\circ$] Frequently, I mostly use use aggregate subreddits, but sometimes I look at individual subreddits.
    \item[$\circ$] Occasionally, I look at aggregate subreddits and individual subreddits about evenly.
    \item[$\circ$] Rarely, I mostly look at individual subreddits.
    \item[$\circ$] Never, I only look at individual subreddits.
\end{itemize}

Do you use reddit more from your computer or from your phone?
\begin{itemize}
    \item[$\circ$] I only use reddit from my computer.
    \item[$\circ$] I mostly use my computer, but sometimes use my phone.
    \item[$\circ$] I use my phone and my computer about evenly.
    \item[$\circ$] I mostly use my phone, but sometimes use my computer.
    \item[$\circ$] I only use reddit from my phone.
\end{itemize}

\subsection*{Subreddit Selection}
In this section, we'll ask you questions specific to the subreddits that you consider yourself a member of.
These subreddits should be subreddits that are important to you,and that you are familiar enough with to feel comfortable commenting on different aspects of how they are run, and how their members interact with one another.

Thank you for selecting subreddits. The next section will ask you about your experiences with each subreddit, individually.

\subsubsection*{Open Ended Subreddit Value Questions}
In this section, we'll ask you questions specifically about your experience in /r/\underline{\hspace{2cm}}.

In your responses, please consider not only the content of /r/\underline{\hspace{2cm}}, but also how it is run, how its members treat one another, and anything else that impacts your experience.

\textit{As it exists right now,} what are a few of the best aspects of the /r/\underline{\hspace{2cm}} community?

\framebox(200,50){}

\textit{If you could change anything,} what are some aspects of the /r/\underline{\hspace{2cm}} community you would like to improve upon?

\framebox(200,50){}

\subsection*{Reflection Questions}
These last two questions ask you to reflect on your experience across all the subreddits you've used.

Generally, what values do you think are important in an online community?
What makes a community healthy?

\framebox(200,50){}

Have you ever stopped participating in any subreddits?
What are some signs that a community isn't worth your time, or isn't a community you want to participate in?

\framebox(200,50){}

}
\newpage

\section{Codebook}\label{app:codebook}













\subsection*{Introduction}
This taxonomy is intended to classify individual idea units, which are focused on a specific aspect of a community.
As such, it is intended to be mutually exclusive - idea units should not be assigned to multiple categories.
If a participants’ response appears to belong to multiple categories, it likely should be subdivided into multiple idea units.

Furthermore, the taxonomy is hierarchical, with the categories being grouped into 9 high level groups. Idea units should be tagged with the most specific category possible - while the high level group is also available as a label itself, it should only be applied when the idea unit in question is either a) too short/vague to be assigned to a more specific category, or b) does not clearly fit into any of the subcategories.

\subsection*{Taxonomy Categories and Descriptions}

\subsubsection*{1) Quality of Content}
Idea units that belong to this category should be commenting on the perceived value/utility of the content in the subreddit, or lack thereof, with exceptions for idea units regarding bullying/offensive content (these belong in category 5a) or idea units regarding the trustworthiness of the content (these belong in category 9b).
As with all categories, idea units in the Quality of Content category should be assigned to as specific of a category as possible.
Thus, many of the idea units that fall into category 1 (as opposed to subcategories 1a-d) are fairly vague/generic.

\underline{Examples}

\textit{I like the content}

\textit{Content is easy to understand}

\underline{Counterexamples}

\textit{The content is funny} belongs in 1a due to the specificity of “funny” (it’s entertaining).

\textit{There are not enough memes} belongs in 1b because the desire for memes is a personal preference.

\textit{Some posts are degrading to women} belongs in 5a because it refers specifically to content this offensive or abusive.

\subsubsection*{1a) Education, Entertainment}
This category should contain idea units that comment on the value of the content, be it educational, entertaining, or some combination (such as “interesting”).
It can sometimes be difficult to distinguish idea units in this category from those in 1b Personal Preferences.
You should use your judgment to determine if the survey respondent’s emphasis is more on the utility of the content (belonging in 1a) the the specific type of content they prefer (belonging in 1b).

\underline{Examples}

\textit{The posts are funny}

\textit{I learn a lot about different mushrooms}

\underline{Counterexamples}

\textit{I especially laugh at the pictures of silly cats} belongs in 1b because the emphasis is on the personal preference towards a specific type of content (silly cat pictures).

\subsubsection*{1b) Personal Preferences}
This category should contain idea units that reflect the respondent’s preferences towards (or against) particular types of content, frequently including idea units on memes, specific subject matter, etc.

\underline{Examples}

\textit{Good memes}

\textit{Custom gaming PCs}

\textit{Less discussion about politics would be good}

\subsubsection*{1c) Curation, Recency, Discovery}
This category focuses on how the content in the subreddit helps the community member discover new things that they may not have otherwise seen, or helps them make sense of large volumes of content (curation).

\underline{Examples}

\textit{/r/seattle helps me keep up to date with news}

\textit{/r/hiphopheads helps me find the best new music}

\subsubsection*{1d) Spam, Reposts, Bots}
Idea units in this category focus on the presence of specific types of content such as spam, reposts, and content submitted by bots.

\underline{Examples}

\textit{There are too many reposts in /r/pics}

\textit{I don’t like the repetitive questions}

\textit{The community bot is helpful}

\textit{Too much spam}

\subsection*{2) Community Engagement}
Idea units in this category focus not on the content, but on the community, either the community as a whole abstract entity, or on specific people or groups of people.
One exception to this is idea units about the credentials or knowledge of people in the community, which belong in 9a Knowledgeable People.
Most idea units in this category are likely to fall into the more specific idea units should be assigned to subcategories 2a and 2b, while very generic or high level idea units should be assigned to this top level category.

\underline{Examples}

\textit{Community}

\underline{Counterexamples}

\textit{Friendly community} belongs in 2a as it comments on a quality (friendliness) of the community as a whole.

\subsubsection*{2a) Quality of Interaction or Community as a Whole}
Idea units in this category comment on both good and bad aspects of the community or of specific interactions within the community.
Note that some specific types of interactions belong in other categories (e.g. size in 4b, harassment in 5a, voting behavior in 8a, rule/norm-breaking in 8b).

\underline{Examples}

\textit{My posts are replied to quickly}

\textit{The community is nice}

\textit{I get to discuss my favorite TV show [in /r/TwinPeaks]}

\subsubsection*{2b) Connection, Universalization}
Idea units in this category should focus on the impact of the community on the survey respondent, such as feeling connected to community members, or aware that there are others out there who feel the same as they do (universalization).

\underline{Examples}

\textit{/r/meow\_irl is so relatable}

\textit{/r/mentalhealth makes me feel connected to others}

\textit{/r/raisedbynarcissists makes me know there are others who feel like me}

\subsubsection*{3) Diversity}
This category focuses on the diversity of content and people within a community.
Almost all idea units in this category should be able to be assigned to one of the two subcategories 3a and 3b.

\subsubsection*{3a) Variety of Content}
This category focuses on the variety of content.
All idea units which praise or critique the diversity of content should fall within this category, with the exception of complaints about reposts (re-submission of the exact same content) which belong in 1d.

\underline{Examples}

\textit{I get to see so many different kinds of snails [on /r/snails]}

\textit{I don’t like how there is a hivemind}

\subsubsection*{3b) Diversity of People}
This category should contain idea units on the diversity of the people within the community or the community as a whole.
Idea units that focus on the diversity (or lack thereof) peoples’ opinions or ideas also belong in this category.

\underline{Examples}

\textit{There are people of so many different backgrounds}

\textit{I don’t like the hivemind about some musicians}

\subsubsection*{4) Size}
Idea units in the size category will most likely fall into either of the two subcategories unless they are very vague.

\underline{Counterexamples}

\textit{Too big} should be assigned to 4b as adjectives “big” and “small” generally refer to the size of the community unless explicitly stated otherwise.

\subsubsection*{4a) Volume of Content}
This category contains idea units that relate to the volume of content (including posts and comments) within a subreddit, or the rate at which this content is posted.

\underline{Examples}

\textit{I wish there were more posts}

\textit{Not enough people post}

\textit{I like that there are always new posts for me to look at}

\textit{Posts get lots of comments}

\textit{Posts (or comments) are submitted so frequently}

\subsubsection*{4b) Size of Community}
This category contains idea units that relate to the size of the community (i.e. the number of people who are in the community or who participate in the community)

\underline{Examples}

\textit{I wish the community were bigger}

\textit{The community is small and close-knit} belongs in 4b as the emphasis is primarily on the size of the community.
Idea units in the category can be similar to those in 2a Quality of Interaction or Community as a Whole.
Use your judgment to determine if the primary emphasis is on the size of the community or on some other qualities

\underline{Counterexamples}

\textit{The community is close-knit} belongs in 2a as the primary emphasis of the idea unit is on the connection within the community, not the size.

\subsubsection*{5) Participation and Inclusion}
Idea units in this category should focus on who participates in the community, and actions and content that explicitly impact who participates and who is included.
Idea units that mention aspects of a community that may have an impact on inclusion, but this impact is not explicitly stated (e.g. `friendly people') should be categorized in 2a.

\underline{Examples}

\textit{Everyone is included}

\underline{Counterexamples}

\textit{They are mean to new members} belongs in subcategory 5a as it relates to behavior which is specifically abusive or harassing

\subsubsection*{5a) Offensive, Abusive, Harassing Content or Behaviors}
Idea units in this category relate to content and behaviors that are offensive, abusive, or harassing to specific people, groups of people, or in general.
You should use your best judgment to interpret if the content is offensive to the respondent, in which case it belongs in this category, or if it is simply not in alignment with their personal preferences, in which case it belongs in 1b.

\underline{Examples}

\textit{I wish there were less jokes that make fun of women} is offensive

\textit{Bullying}

\textit{[/r/pics] pictures of male genitalia} belongs in this category because in the context of /r/pics, pictures of genitalia are unexpected and therefore likely offensive.

\underline{Counterexamples}

\textit{[/r/nudes] pictures of male genitalia} does not belong in this category because the most likely interpretation of “pictures of male genitalia” in the context of /r/nudes is that of a personal preference (given that such pictures would be expected on the /r/nudes subreddit) and therefore this idea unit is best categorized under 1b.

\textit{People are quick to criticize} belongs in 5 because criticism isn’t inherently offensive

\subsubsection*{5b) Outsiders and Demographics}
Idea units in this category should explicitly relate to who participates in the community, and their perceived out/in-group status.

\underline{Examples}

\textit{[/r/wichita] too many people from outside the city post here}

\textit{Posts are mostly submitted by people from outside the community}

\subsubsection*{5c) Tools for Participation}
This category relates to tools (such as wikis, FAQs, and stickied posts) that help improve participation.

\underline{Examples}

\textit{The bot that welcomes new members is really nice}

\textit{There should be be an FAQ explaining how to post}

\subsubsection*{6) Technical Features}
Idea units in this category and subcategories should relate to technical features offered (or not offered by reddit).
As with all categories, the most specific subcategory should be used.
As such, this top-level category should only be used for broad idea units that do not fit into 6a-c.

\underline{Examples}

\textit{Videos don’t load sometimes}

\textit{I prefer old reddit to the redesign}

\textit{The mobile app works well}

\subsubsection*{6a) Flairs, Tags, NSFW Labels}
Idea units in this category should relate to flairs (small icons or text strings associated with usernames), tags applied to posts (this includes stickied/pinned posts or comments), and NSFW (Not Safe for Work) labels.

\underline{Examples}

\textit{[/r/colonoscopy] NSFW tags are really well used}

\textit{Flairs are helpful for knowing who is who}

\textit{Post tags for categories are convenient}

\textit{Sometimes NSFW content isn’t marked as such}

\subsubsection*{6b) Search, Filters}
Idea units in this category relate to searching for content or filtering for specific types of content.

\underline{Examples}

\textit{I like that I can easily search for pictures}

\textit{I wish I could filter by experience level}

\textit{It would be nice to be hide NSFW content} belongs in this category because it relates primarily to the ability to filter out NSFW content, not the NSFW label itself.

\subsubsection*{6c) Recommendation Systems}
Idea units in this category relate to recommendation systems for finding similar content.

\underline{Examples}

\textit{I wish there were recommendations for seeing more posts like the ones I like}

\subsubsection*{7) Moderation and Moderators}
This category contains idea units regarding who the moderators are, and how they perform their job (or fail to perform their job).

\noindent
On reddit, moderators are responsible for:
\begin{itemize}
    \item Setting rules
    \item Enforcing rules, including
    \item Removing posts
    \item Banning users
    \item Updating tags and flairs and NSFW labels
    \item Recruiting and selecting new moderators
    \item Communicating their work to the community
\end{itemize}

\noindent
Note that this category should only contain idea units relating specifically to moderators and their actions.
Idea units regarding rules themselves (e.g. requests for new rules) should go in 8 and its subcategories.

\underline{Examples}

\textit{Moderators are very friendly and reply quickly to messages}

\textit{The mods are corrupt and on a power-trip}

\textit{I wish the community was involved in new rules being created}

\textit{Rules aren’t well enforced}

\textit{Moderators enforce rules unfairly}

\underline{Counterexamples}

\textit{I wish content from 4chan was banned} belongs in 8b because it is a request for a rule.

\subsubsection*{8) Norms and Rules}
Idea units in this category and its subcategories should relate to rules and norms, explicitly stated or not. Idea units relating to how rules are enforced, or other aspects of moderation, belong in 7 Moderators.

\subsubsection*{8a) Voting Behavior}
Idea units in this category should relate specifically and explicitly to voting behavior (upvotes and downvotes on reddit).

\underline{Examples}

\textit{People shouldn’t use downvote as a disagree button}

\underline{Counterexamples}

\textit{Upvotes help me find the best music} belongs in 1c because it is primarily relating the the curation mechanism provided by votes, not how people vote.

\subsubsection*{8b) Adherence to Norms and Rules}
Idea units in this category should relate to the subreddits’ rules and norms and how the community adheres to norms and rules.
Many idea units in this category will be specific to the type of content within the subreddit, so checking the subreddit the idea unit corresponds to may be helpful.
This category should also include requests for rules.
Implicit norms can be particularly complex and context dependent, so consider these carefully and use your best judgment.
Note that idea units relating to the enforcement of rules belong in 7 Moderators.

\underline{Examples}

\textit{It is frustrating when newbs ask questions that are already answered in the FAQ}

\textit{[/r/lgbtq] People sometimes ask us to basically pick a label for them - this is a deeply personal identity, not something we can pick for other people}

\textit{I wish content from 9gag was banned}

\textit{They should allow memes}

\textit{[/r/birdspotting] People should be required to include the location the saw the bird in their request for ID}

\textit{[/r/ElderScrolls] People should specify which version of the game they’re referring to}

\underline{Counterexamples}

\textit{Rules aren’t enforced strictly enough} belongs in 7 because it primarily relates to enforcement, which is performed by moderators.

\subsubsection*{9) Trust}
Idea units in this category relate to the trustworthiness (or lack thereof) of the people and the content within subreddits.
The top-level category should be used only when it is not possible to differentiate between idea units commenting on the people (9a) and the content (9b).

\underline{Examples}

\textit{Students post their honest experiences and opinions so it’s legit} belongs in the top-level category because it is commenting both on the people (students are honest) and the content (posts are honest)

\subsubsection*{9a) Knowledgeable People}
Idea units in this category should relate to the knowledge, trustworthiness, or credentials of the people in the community.

\underline{Examples}

\textit{There are lots of doctors so I know what they’re saying is correct}

\textit{People are mostly honest}

\subsubsection*{9b) Trustworthy Content}
Idea units in this category should relate to the veracity or trustworthiness of content in the community.

\underline{Examples}

\textit{There is no fake news}

\textit{[/r/whatisthisplant] Multiple people often confirm the same ID so I know it’s right}

\subsubsection*{10) Exclude/Unintelligible/Off-topic}

\subsection*{Notes for Raters}
Responses are confidential. Please do not disclose them.
Use your best judgment, and refer back to this codebook as necessary. Feel free to use the comments column to note any responses you were especially uncertain of.
When labeling, mark the number and letter of the category in the label column. For example, an idea unit that you classify into ``Tools for Participation'' should be labeled ``5c''.
Don’t worry if the labels are deeply skewed - in our initial sample, categories 1a, 1b, and 2a were by far the most common.
Leave blank idea units that you think are not categorizable, and add a comment explaining why.

\end{document}